\documentclass[twocolumn,showpacs,preprintnumbers,amsmath,amssymb]{revtex4}

\usepackage{subfigure}
\usepackage{graphicx}
\usepackage{dcolumn}
\usepackage{bm}

\begin{document}


\title{New set of measures to analyze dynamics of non-equilibrium structures}

\author{Girish Nathan}
\email{Girish.Nathan@mail.uh.edu}
\author{Gemunu Gunaratne}%
 \email{Gemunu@mail.uh.edu}
\affiliation{%
Department of Physics, University of Houston, 4800 Calhoun Blvd., Houston, TX 77204,USA\\
}%

\date{\today}

\begin{abstract}
We introduce a set of statistical measures that can be used to quantify non-equilibrium surface growth. They are 
used to deduce new information about spatiotemporal dynamics of model systems for spinodal decomposition and surface deposition. Patterns growth in the
Cahn-Hilliard Equation (used to model spinodal decomposition) are shown to exhibit three distinct stages. Two models of surface growth, namely the 
continuous Kardar-Parisi-Zhang (KPZ) model and the discrete Restricted-Solid-On-Solid (RSOS) model are shown to have different saturation exponents. 
\end{abstract}

\maketitle

In recent years, there has been a considerable amount of effort to analyze non-equilibrium interfacial growth and 
pattern formation in experimental and model systems \cite{bray,kpz,kk,hoh,langer,famvic,hhzhang,barabasi,dassarma1,dassarma2}. The phenomena studied include spinodal decomposition \cite{bray,langer,tor1,tor2}, chemical pattern formation \cite{hoh},
surface growth \cite{barabasi,kpz,kk,hhzhang} and epitaxial growth \cite{dassarma1,dassarma2}. Microscopic modeling of these phenomena is highly complex, and 
most microscopic details of such structures depend on initial conditions and stochastic effects. Hence model systems are often used to extract statistical properties of these structures and to determine the physical processes that are most relevant for their growth. Some of the model systems that have been introduced to study the spatio-temporal dynamics 
of the aforementioned systems are the Cahn-Hilliard Equation (CHE) \cite{bray}, the Kardar-Parisi-Zhang (KPZ) model \cite{kpz}, the Restricted-Solid-On-Solid (RSOS) model \cite{kk}, and the
Swift-Hohenberg Equation (SHE) \cite{hoh,swifthoh}.

In order to confirm if a given model can accurately represent a pattern forming process, it is necessary to compare as many statistical measures as 
possible. Such a comparison between model systems and physical systems is also needed to validate claims of universality. Unfortunately, there are only a
handful of measures that are available to be used for such comparisons. The aim of this work is to introduce a new family of such measures. 

Commonly used statistical measures to analyze surface growth and patterns include surface roughness (i.e., the standard deviation of the heights)
 $W_{L}(t)$, where $L$ is the lattice size and $t$ the time, the correlation length \cite{barabasi}, and the domain size \cite{tor1,tor2}. For the KPZ and RSOS interfaces, 
$W_{L}(t) \sim t^{\beta}$ when $t \ll L^{z}$. At very large times, $W_{L}(t \rightarrow \infty) \sim L^{\alpha}$ \cite{barabasi}.  
The growth exponent $\beta$, the dynamic exponent $z$, and the roughness exponent $\alpha$, depend only on the dimensionality of the growth process and are independent of $L$ apart from finite-size
scaling corrections \cite{famvic}. It has been found numerically that these exponents are the same in all dimensions for the KPZ and the RSOS models. Based
on the results, it has been suggested that KPZ and RSOS belong to the same universality class. This is a non-trivial statement, since although both models
consider the competition between random deposition and diffusion, RSOS is a discrete counterpart of the KPZ with distinct microscopic dynamics. The 
availability of additional statistical measures can be used to test the validity of the assertion of universality. In fact, one of our conclusions, based on
the new measures, is that there are statistical features that are not common to KPZ and RSOS models.

Non-equilibrium pattern formation and dynamics has also been extensively studied in the context of spinodal decomposition using the Cahn-Hilliard Equation \cite{bray,tor1,tor2}. 
The spatio-temporal dynamics can be classified into two temporal regimes - an early stage where there are many small clusters (cluster size $\ll$ $L$) and a late stage where there are large 
interconnected domains and cluster sizes are comparable to L. The domain size in the late phase is found to grow in time with an exponent $\frac {1} {3}$ \cite{tor2}.

Consider a planar pattern represented by a scalar field $U(\vec{x},t)$ which can be the height of an interface, or some relevant intensity field. One feature
not captured by the correlation length and roughness is the amount of curvature of the contour lines of $U$. The possibility of using curvature for such an
analysis has been proposed before \cite{hu,gem1,gem2,gem3}, but in practice the results are very sensitive to noise. The underlying reason is that the 
evaluation of  
$\kappa=(U_{xx} U_{y}^{2}+U_{yy} U_{x}^{2}-2 U_{xy} U_{x} U_{y})/(U_{x}^{2}+U_{y}^{2})^{\frac{3}{2}}$ is very sensitive to errors in the denominator, when it is small. In place of $\kappa$, one can use another measure, namely the determinant of the
Hessian normalized by the variance of $U$, $\Delta=(U_{xx} U_{yy}-U_{xy}^{2})/Var(U)$. Unlike $\kappa$, the calculation of $\Delta$ is fairly insensitive
to noise. Further, for typical local structures, $\Delta$ is proportional to $\kappa$. For example if $U(r)=e^{- \alpha r}$, $\Delta=-\alpha^{3} e^{-2 \alpha r} \kappa$; for $U(r)=sin(kr)$, 
$\Delta=-k^{3} sin(2kr) \kappa/2$.

The new measures we define are,
\begin{equation}
\mu(\delta,t)={\left (\frac{\int d^{2} \vec{x}  {|\Delta|}^{\delta} } {\int d^{2} \vec{x}} \right )}^{1/4 \delta}.
\end{equation}
Note that for each $\delta$, $\mu(\delta,t)$ has dimensions of inverse length. Furthermore, the measures $\mu(\delta,t)$ are invariant under all rigid Euclidean transformations (i.e., translations, rotations, and reflections) of a pattern.
The use of moments, $\delta$, allows us to emphasize regions with different values of $\Delta$, thereby providing an array of lengthscales associated with the  
structure. The use of multiple lengthscales is similar in spirit to using multifractals \cite{hal,hen1} to characterize strange attractors, although the spatio-temporal nature of the dynamics for the models
considered in this Letter makes a further connection difficult.
 In our analysis, we define $\sigma(\delta,t)$ as the growth rate of $\mu(\delta,t)$, i.e., $\sigma(\delta,t) \equiv log(\mu(\delta,t))/log(t)$.

The organization of the paper is as follows. We first discuss our results of our analysis for the CHE, and describe distinct 
stages in the spatio-temporal dynamics using $\mu(\delta,t)$. We then proceed to analyze the KPZ and RSOS models and show 
that during the initial stage, $\mu(\delta,t)$ does indeed lend additional credence to the suggestion that in the strong non-linear coupling limit,
they do belong in the same universality class. However, our analysis of the late stage shows that the  
saturation exponents for the two models are different. The computational techniques used to obtain the data for these models is discussed in 
detail elsewhere \cite{gem3} and will not be expanded upon here. 
  
The CHE models spinodal decomposition using the dynamics of a conservative field $\psi(\vec{x},t)$ via 
\begin{equation}
\frac{\partial \psi}{\partial t}=\frac{1}{2} {\nabla}^{2}(-{\nabla}^{2} \psi-\psi+{\psi}^{3})+\sqrt \epsilon \eta(\vec{x},t).
\end{equation}  
Here, $\eta$ is delta-correlated noise of zero mean and amplitude 1.
The Crank-Nicholson method \cite{nr} used to integrate Equation (2) allows us to choose timesteps as large as $0.4$ and enables us to investigate the dynamics of the CHE to times large enough to observe saturation of $\mu(\delta,t)$, e.g., 
$T=800,000$ for the lattice size $L=32$ and $T=1,600,000$ for the lattice size $L=128$. To obtain $\mu(\delta,t)$, we used the averages of 32 runs for $L=32$, 16 runs for $L=64$, and
8 runs for $L=128$. The errorbars on all curves are calculated by averaging over these realizations. 

The dynamics of domain growth is as follows: Beginning from a random initial configuration, $|\psi(\vec{x},t)|$ and the domain size 
grow in time. For sufficiently large $t$, $\psi(\vec{x},t)$ reaches its 
equilibrium values of $-1$ or $+1$ \cite{tor2,bray}. 
\begin{figure}
\begin{center}
\includegraphics[scale=0.4,angle=0]{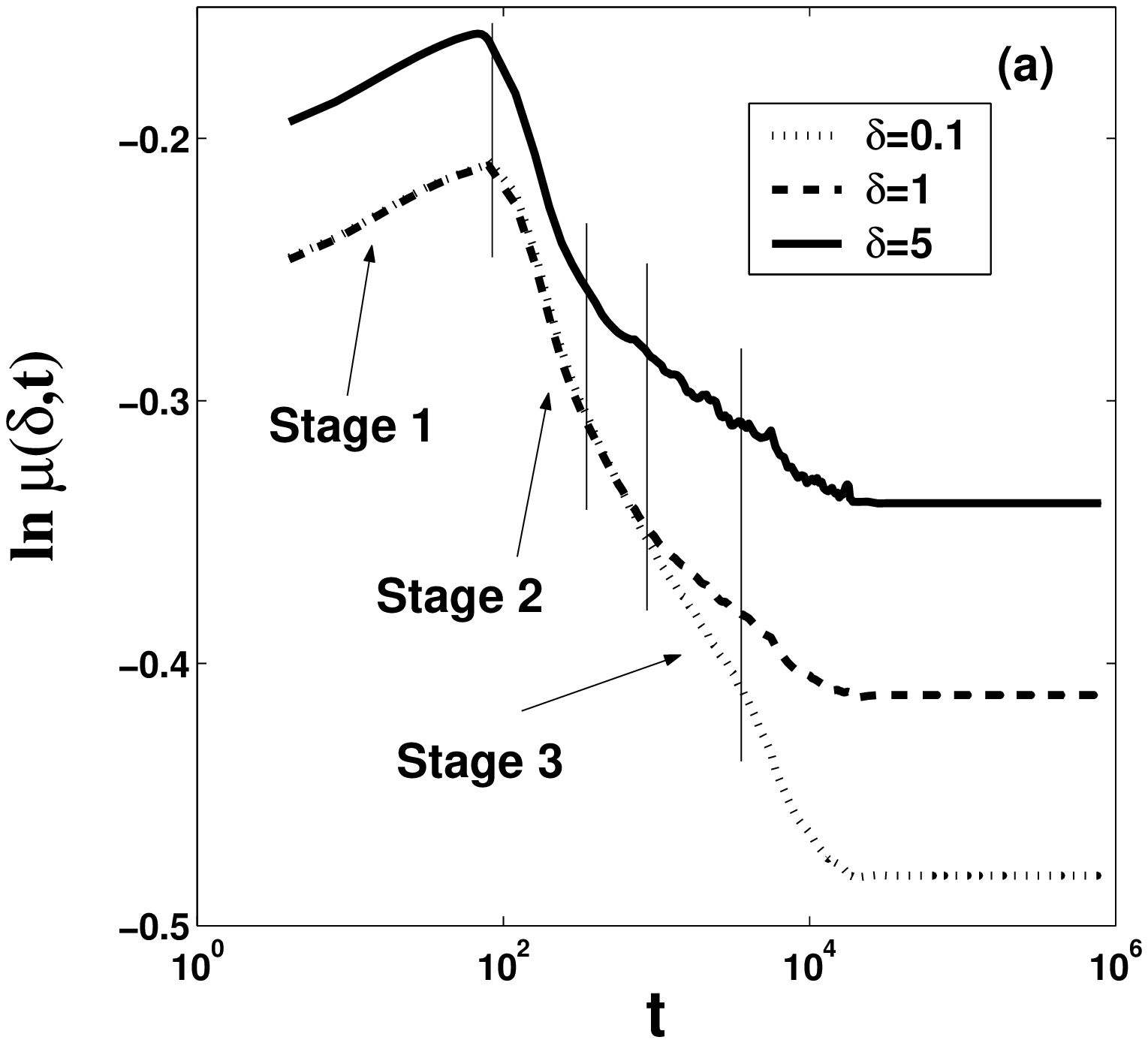}
\end{center}
\end{figure}
\begin{figure}
\begin{center}
\includegraphics[scale=0.4,angle=0]{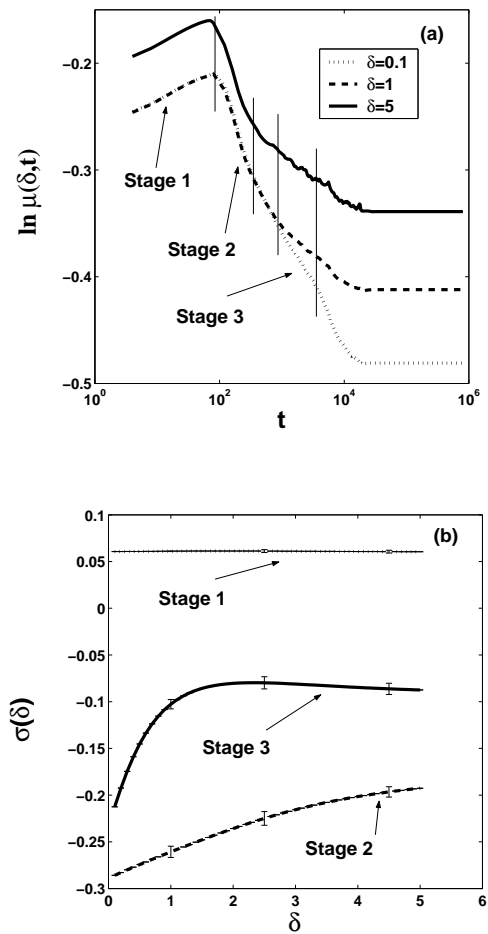}
\caption[]{Plots of the measure in time for various moments. $L=32$. (a) The noise-free case for different $\delta$, (b) The slopes versus $\delta$ for Stage 1,2, and 3 for the noise-free case. \label{f1}}
\end{center}
\end{figure}
The behavior of $\mu(\delta,t)$ is shown in Fig. 1(a). 
The dynamics can be separated into three stages (marked Stage 1,2, and 3 in the figure). Initially, $\mu(\delta,t)$ appears to grow as a power law in time for about one and a half decades, as seen
in Fig. \ref{f1}(a). There is a formation of small domains from the random initial field. As seen from the topmost curve of Fig. \ref{f1}(b), the growth rates $\sigma_{1} (\delta,t)$ for all $\delta$ are nearly identical and independent of $t$
. During this stage, one contribution to $\mu(\delta,t)$ comes from points near the domain boundaries. 
This is seen by plotting the height along any row or column. 
Although the formation of domains reduces the number of interfacial points
, this decrease is countered by the fact that since $\psi(\vec{x},t)$ has not 
reached its equilibrium values, the internal points of a domain provide an increasing contribution to the integrand. Consequently,
$\mu(\delta,t)$ increases during this stage. 

The crossover between Stages 1 and 2 occurs when the field $\psi$ saturates to its equilibrium
values. A histogram of interfacial heights clearly shows the concentration near 
the equilibrium values of $\pm 1$. Beyond this stage, interior points of a domain no longer contribute to $\mu(\delta,t)$.
As a result, the aforementioned decrease in interfacial boundary  points now leads to a decrease in   
$\mu(\delta,t)$.  
The slopes $\sigma_{2} (\delta,t)$ in this stage are plotted in the lower curve of Fig. \ref{f1}(b). We find that in this stage, the
distinct moments relax at different rates (i.e, $\sigma_{2} (\delta,t)$ is $\delta$-dependent). Moreover, it is seen from Fig. \ref{f1}(b) that
the rate of relaxation of the larger moments is smaller than that for the lower moments, implying that sharper features (those with 
larger curvature) change 
relatively slowly. For example, straightening of a domain boundary wall occurs at a faster rate than elimination of a small domain. 
 Running averages reveal that $\sigma_{2}(\delta,t)$ is uniform until the end of this stage. Observe one critical difference
between Stages 1 and 2, namely that the difference in relaxation of the distinct lengthscales cannot be determined without an array of measures. 
\begin{figure}
\begin{center}
\includegraphics[scale=0.4,angle=-90]{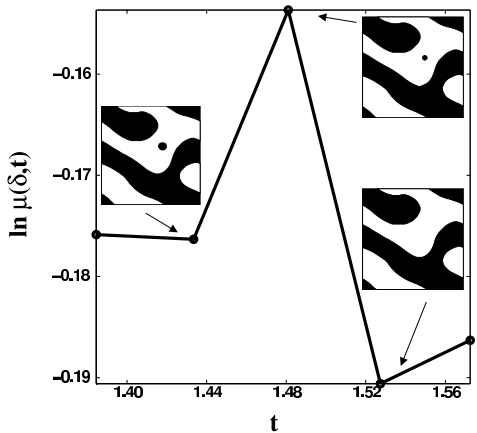}
\caption[]{Bubble disappearance for the $L=128$ lattice. The actual time is obtained by scaling the x-axis values by a factor $10^{5}$. Notice how the peak in the measure corresponds to the
bubble disappearing at approximately time $148,000$. This is for $\delta=3.4$. \label{f2}}
\end{center}
\end{figure}

The domain size as measured from the two point correlation function \cite{tor2} becomes comparable to $L$ between Stages 2 and 3 and $\sigma_{2} (\delta,t)$ changes during the transition. 
The spatio-temporal dynamics beyond this involves a coarsening of the large domains and elimination of the small ones. The disappearance of 
small domains can be identified by peaks in $\mu(\delta,t)$ for large values of $\delta$. These peaks result from the large curvature of a 
disappearing domain. Fig. \ref{f2} shows the effect of such an event on $\mu(\delta,t)$. It is clear that the 
elimination of a domain is accompanied by a peak in $\mu(\delta,t)$.
The decay rates $\sigma_{3} (\delta,t)$ for the third stage are not uniform in time, unlike the previous stages. An example is shown in the central curve of 
Fig. \ref{f1}(b). Although $\sigma_{3} (\delta,t)$ varies in time, the general form is the same; there is a $\delta$-dependence for small values of 
$\delta$ but a saturation for higher values. 
At the end of Stage 3, only a handful of large domains remain; the only dynamics beyond this is an extremely slow straightening of the
interfaces.   

Surface roughness and domain size cannot provide such detailed information on the spatio-temporal dynamics. For a given structure, $W_{L}(t)$
saturates at the end of Stage 1, and there is no discernible difference beyond this. The domain size grows as $R(t) \sim t^{\frac{1}{3}}$ at large times but 
also saturates between Stage 2 and 3, and gives no further information.

We have also investigated the effect of adding  zero-mean external noise to the CHE. We find that in the presence of noise,
the dynamics in Stage 3 is faster than in the noise-free case, i.e., $|\sigma_{3} (\delta,t)|$ is larger for the noisy dynamics. 
We also find that the saturation time of $\mu(\delta,t)$ is proportional to $\sqrt \epsilon$.

The KPZ equation is a paradigmatic model of non-equilibrium interfacial growth in the presence of lateral
correlations \cite{kpz,hhzhang}. It gives the local growth of the height profile $h(\vec{x},t)$ at substrate position $\vec{x}$ and 
time $t$ as
\begin{equation}
\frac{\partial h(x,t)}{\partial t}=\nu {\nabla}^{2} h +{\lambda} {(\nabla h)}^{2}+\eta(\vec{x},t).
\end{equation}
Here $\nu$ is the diffusion parameter which smoothens the interface and $\lambda$ the prefactor of the nonlinear term which tends to amplify large slopes. $\eta({\vec{x},t})$ is delta-correlated noise of zero mean which represents a random particle flux.
RSOS is a discrete counterpart of the KPZ. Here, particles are added to a randomly chosen site $i$ if and only if the 
addition ensures that all nearest-neighbor height differences $|\Delta h| \leq \Gamma$, where $\Gamma$ is some pre-determined positive 
integer. 

Roughness and correlation function analyses of the dynamics of both KPZ and RSOS models show that 
their growth and roughness exponents take similar values in all dimensions, i.e., they are in the same universality class for large values of $\lambda$. 
\begin{figure}
\begin{center}
\includegraphics[scale=0.4,angle=0]{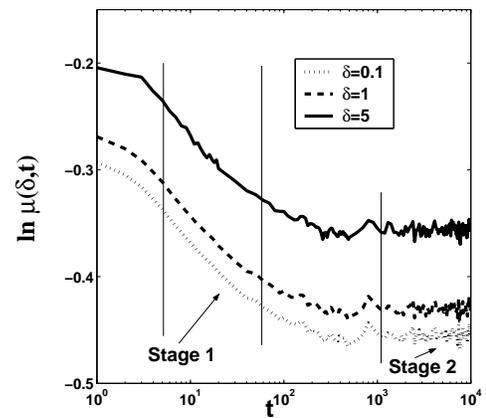}
\caption[]{ Measure for the KPZ model for different $\delta$. $L=128.$ \label{f3}}
\end{center}
\end{figure}
\begin{figure}
\begin{center}
\includegraphics[scale=0.4,angle=0]{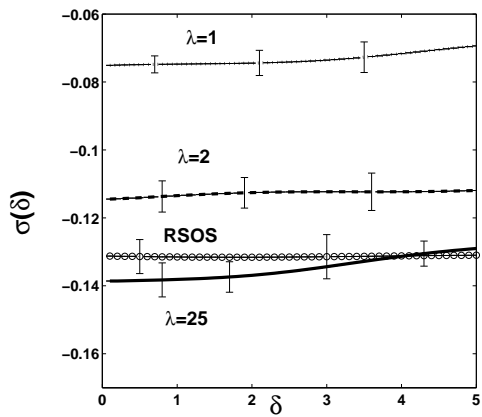}
\caption[]{ The slopes vs. $\delta$ for various $\lambda$ values compared with the RSOS slopes. \label{f4}}
\end{center}
\end{figure}
In the initial stages, the correlations spread across the entire lattice and the correlation length $\zeta \sim t^{\frac{1}{z}}$, where
$z \sim 1.6$. It has been established numerically \cite{amarfamily1} that the strong non-linear coupling
limit of KPZ corresponds to the RSOS model in terms of the growth and the saturation exponents of the surface
roughness.

The behavior of $\mu(\delta,t)$ in time for the KPZ model ($L=128$) is shown in Fig. \ref{f3}. The dynamics is qualitatively
similar to the behavior of the surface roughness, and can be divided into two distinct stages. 
In Stage 1, the lateral correlations spread across the entire lattice and we find 
that $\mu(\delta,t)$ decreases as $t^{-\sigma_{1}(\delta,t)}$ where $\sigma_{1}(\delta,t)$ is independent of $\delta$ and $t$
within numerical errors. The plots labelled by the dotted, dashed and solid lines in Fig. \ref{f4} correspond to $\sigma_{1}(\delta,t)$ 
for three values of $\lambda$ of the KPZ model ($\lambda=1$, $\lambda=2$, and $\lambda=25$ respectively) 
during the first stage. $\mu(\delta,t)$  for the RSOS model exhibits a similar behavior during Stage 1.
In Fig. \ref{f4}, the plot with circles shows $\sigma_{1}(\delta,t)$ versus $\delta$ for the RSOS model. 
This corroborates the assertion that for the large $\lambda$ limit, KPZ corresponds to the 
RSOS models in terms of the growth exponents, since it is seen clearly that $\sigma_{1}(\delta,t)$ for KPZ with  
$\lambda=25$  and for RSOS are within the errorbars. 

In Stage 2 of Fig.\ref{f3}, $\mu(\delta,t)$ saturates to a $L$-dependent value (given by $L^{\gamma(\delta)}$)
where $\gamma(\delta)$ is the saturation exponent. However, in this case,
we find that $\gamma(\delta)$ for KPZ and RSOS models, even at strong coupling, are very different
. For example, $\gamma(\delta)$ for the RSOS model is found
to be $\gamma(0.1)=-0.095 \pm 0.004$ while that for the KPZ model ($\lambda=25$) is $\gamma(0.1)=-0.075 \pm 0.007$ when $\delta=0.1$. 
For $\delta=1.0$, the saturation exponent for RSOS is $\gamma(1.0)=-0.083 \pm 0.002$ while that for the KPZ is
$\gamma(1.0)=-0.068 \pm 0.007$.  

In summary, we have introduced a set of characteristics $\mu(\delta,t)$  that can be used to study spatio-temporal dynamics
of systems represented by a scalar field $U(\vec{x},t)$. At a given instant, $\mu(\delta,t)$ are a set of inverse lengthscales associated with a pattern, and are defined
in terms of the determinant of the Hessian. Large values of $\delta$ emphasize regions with larger curvature of contour lines. 

The availability of a family of indices allows us to provide a more comprehensive description than is possible from individual measures. For example, it 
was shown for the CHE that during Stage 2, relaxation of the distinct lengthscales associated with the structure occurs at different rates. This is very
different from Stage 1 where all such scales grow as $\mu(\delta,t) \sim t^{0.06}$. We are also able to identify instances where small domains of the
pattern disappear by searching for peaks in $\mu(\delta,t)$ for large values of $\delta$.

Analysis of the KPZ and the RSOS models shows two stages in pattern evolution. For KPZ, $\mu(\delta,t)$ relaxes at a rate that depends on $\lambda$ 
during stage 1. The rate of decay $\sigma_{1}(\delta,t)$ for large values of $\lambda$ (typically $\lambda \geq 15$) is seen to be the same as for the RSOS model,
thus reinforcing previous claims that both models belong to the same universality class. However, our analysis provides an additional piece of information, namely that all lengthscales $\mu^{-1}(\delta,t)$ associated with the spatio-temporal dynamics of these interfaces decay at the same rate. $\mu(\delta,t)$ saturates in stage 2 for both models, and the
saturation value depends on the system size $L$ as $L^{\gamma(\delta)}$. Interestingly, the function $\gamma(\delta)$ for the two models is different;
after saturation, the contours of the two models exhibit non-universal characteristics. It is thus possible to determine which of these models better represents the growth
of an experimental interface.

The authors thank R. Rajesh and K. E. Bassler for a critical reading of the manuscript and for useful discussions. This research is partially funded by a grant from the National Science Foundation.

\bibliography{hessian-new}
\end{document}